\documentclass[aps,prb,twocolumn,superscriptaddress,showpacs]{revtex4-1}
\usepackage{graphicx}  
\usepackage{dcolumn}   
\usepackage{bm}        
\usepackage{amsfonts,amssymb,amsbsy,amsmath,amsthm}
\usepackage{setspace}
\newcommand{\Gu}{u}
\def\mathclap#1{\text{\hbox to 12pt{\hss$\mathsurround=15pt#1$\hss}}}

\begin{document}

\title{Moments of work  in the two-point measurement protocol for a driven open quantum system}
\author{S. Suomela} \email{samu.suomela@aalto.fi}
\affiliation{Department of Applied Physics and COMP Center of Excellence, Aalto University School of Science, P.O. Box 11100, FI-00076 Aalto, Finland}

\author{P. Solinas} 
\affiliation{SPIN-CNR, Via Dodecaneso 33, I-16146 Genova, Italy}

\author{J. P. Pekola}
\affiliation{Low Temperature Laboratory (OVLL), Aalto University School of Science, P.O.
Box 13500, 00076 Aalto, Finland}

\author{J. Ankerhold} 
\affiliation{Institut f\"{u}r Theoretische Physik, Universit\"{a}t Ulm, Albert-Einstein-Allee 11, 89069 Ulm, Germany}

\author{T. Ala-Nissila}
\affiliation{Department of Applied Physics and COMP Center of Excellence, Aalto University School of Science, P.O. Box 11100, FI-00076 Aalto, Finland}
\affiliation{Department of Physics, Brown University, Providence, Rhode Island 02912-1843, USA}


\date{September 30, 2014}

\begin{abstract}
We study the distribution of work induced by the two-point measurement protocol for a driven open quantum system.
We first derive a general form for the generating function of work for the total system, bearing in mind that the Hamiltonian does not necessarily commute with its time derivative. Using this result we then
study the first few moments of work by using the master equation of the reduced system, 
invoking approximations similar to the ones made in the microscopic derivation of the reduced density matrix. 
Our results show that, already in the third moment of work, correction terms appear that involve commutators between
the Hamiltonian and its time derivative. To demonstrate the importance of these terms, we consider a sinusoidally, weakly
driven and weakly coupled open two-level quantum system, and indeed find that already in the third moment of work
the correction terms are significant.  We also compare our results
to those obtained with the quantum jump method and find a good agreement. 
\end{abstract}

\maketitle

\section{Introduction}

For microscopic systems driven out of equilibrium the fluctuation theorems, e.g., Refs. \onlinecite{Bochkov1977,Jarzynski1997,Crooks1999, Seifert2005},
provide a powerful tool to analyze the thermodynamic nature of non-equilibrium processes beyond the linear response regime.
When the microscopic system can be described in terms of classical mechanics, the fluctuation theorems have been examined
for several systems \cite{Liphardt2002,Harris2007,Greenleaf2008,Imparato2008,Saira2012,Kutvonen2013, Koski2013}. However, when described in terms of quantum mechanics, the situation is
more problematic. In quantum systems, it is far from obvious how to treat certain thermodynamical quantities such as work $W$ that relate to the physical path of the system rather than to the state (wave function).

Work appears in the classical Jarzynski equation (JE) $\langle e^{-W / k_B T}\rangle = e^{- \Delta F / k_B T}$, where $\Delta F$ is the free-energy difference between the initial equilibrium and the final states, and
the brackets $\langle \cdot \rangle$ denote averaging over an infinite number of repetitions. Trying to generalize the JE to the quantum regime
has caused much debate about how to define $W$ in a physically meaningful way.
Earlier quantum treatments of the JE were based on defining a genuine work operator \cite{Yukawa2000, Chernyak2004,Allahverdyan2005,Engel2007}. Yet since work is not a traditional quantum observable \cite{Talkner2007}, the use of a quantum work operator leads to corrections to the JE. It can be recovered by another approach,
known as the two-point measurement protocol \cite{Kurchan2000,Mukamel2003,Monnai2005,Campisi20111}, in which the energy of the closed system is measured at the
beginning and at the end of the process and there is no dissipated heat. The work of a single trajectory is then defined as the energy difference of the final and initial measurement outcomes. In the case of open systems assuming that the interaction Hamiltonian is negligible, the energy measurement of the total
closed system can be approximated by measuring the energy of the reduced system and the environment separately.

In a recent paper \cite{Hekking2013} the quantum jump method, also known as the Monte Carlo wave function method (MCWF), was proposed as an efficient way to 
discuss the problem of determining the statistics of work in driven quantum systems with dissipation. By interpreting a jump between the eigenstates of the Hamiltonian as an emission and absorption of a 
photon to the heat bath, the total energy exchanged between the system and the heat bath due to the jumps is then interpreted as heat. The work can then be 
defined as the energy difference between the initial and final states of the system plus the heat released to the heat bath. 
It should be noted that with this definition a possible energetic contribution from the interaction between the system and the heat bath 
was not taken into account in work \cite{Solinas2013,Salmilehto2014}.

In this paper, we analyze in detail the first few moments of work by using the master equation approach for an open quantum system. To characterize the stochastic nature of $W$ and its
distribution, it is natural to consider the moments of work instead of directly trying to  calculate exponential averages such as that
in the JE, which is a formidable
task for open quantum systems in general. The first moment gives the mean work done, the second moment gives the variance and 
the third moment gives the skewness of the work distribution for non-Gaussian distributions.  As the first step we derive the 
two-point measurement protocol generating function without making the implicit assumption in Ref.  \onlinecite{Esposito2009} that 
the total system Hamiltonian commutes with its time derivative. This result allows us to derive general expressions for the first three
moments of work, which we compare with results obtained using the generating function of Ref.  \onlinecite{Esposito2009} (Eqs. (17),(18),(22) and (23) in Ref.  \onlinecite{Esposito2009}). Our results show
that only the first two moments of work are identical in the two approaches above, and nontrivial correction terms appear to the third and higher moments when the
Hamiltonian does not commute with its time derivative. To study this issue in a specific case we consider the weakly coupled and
weakly driven open two-level quantum
system of Ref. \onlinecite{Hekking2013}, where we invoke approximations similar to those used in the microscopic derivation of the 
Lindblad equation of the reduced system. The test system describes, for instance, a Cooper box coupled capacitively or a dc superconducting quantum interference device (dc-SQUID) coupled inductively to a calorimeter \cite{Pekola2013}. When calculating the dynamics of the test system, we neglect the interaction Hamiltonian in the energy measurements. We indeed find that our results for the first three moments are in agreement with the quantum jump results. When comparing the two different generating functions, we find a significant difference in the values of the third moment.

The general results derived here are not restricted to a Cooper box and a dc-SQUID, but can be used for various kinds of superconducting qubits\cite{Clarke2008} and quantum dot circuits\cite{Kouwenhoven1997,Nakamura2010, Kung2012}.

\section{Generating function and moments for work}

In the two-measurement protocol for a closed quantum system, the probability to measure energy $E_0$ at time $t=0$ and $E_{\tau}$ at $t=\tau$ is of the form \cite{footnote}
\begin{equation}
P[E_\tau,E_0]=\text{Tr}\lbrace \hat{P}_{E_\tau} \hat{U}(\tau,0) \hat{P}_{E_0} \hat{\rho}_0 \hat{P}_{E_0} \hat{U}^\dagger(\tau,0) \hat{P}_{E_\tau}  \rbrace, \label{Eq:prob}
\end{equation}
where $\hat{U}(\tau,0)=\mathcal{T}_\leftarrow \exp{\left(-\frac{i}{\hbar} \int_0^\tau dt \hat{H}(t)  \right) }$ is the unitary time evolution operator, $\mathcal{T}_\leftarrow$ describes the chronological time ordering and the projection operators are given by $\hat{P}_{E_t}=| E_t \rangle \langle E_t |$, where $| E_t \rangle$ is the state corresponding to the measurement result $E_{t}$ at time $t$. 
The corresponding generating function is the Fourier transform of  $P[E_\tau,E_0]$ \cite{Esposito2009} (the calculation is also given in Appendix \ref{A}):
\begin{eqnarray}
G(\Gu)&=&\sum\limits_{E_0,E_\tau} e^{i \Gu (E_\tau-E_0)} P[E_\tau,E_0]  \nonumber \\
	 &=& \text{Tr} \left\lbrace \hat{U}_{\Gu /2}(\tau,0)  \bar{\hat{\rho}}_0 \hat{U}_{-\Gu/2}^\dagger(\tau,0)  \right\rbrace,  \label{Eq:Gene}
\end{eqnarray}
where
\begin{eqnarray}
\hat{U}_{\Gu}(\tau,0)&=&e^{i \Gu \hat{H}(\tau)} \hat{U}(\tau,0)  e^{-i \Gu \hat{H}(0)}; \label{Eq:Ulambda} \\ \bar{\hat{\rho}}_0&=&\sum_{E_0} \hat{P}_{E_0}\hat{\rho}_0 \hat{P}_{E_0},
\end{eqnarray}
 and $\hat{\rho}_0$ is the initial density matrix. If the initial density matrix is diagonal in the first measurement basis, then $\bar{\hat{\rho}}_0=\hat{\rho}_0$. 

The differentiation of the evolution operator $\hat{U}_{\Gu}(\tau,0)$ [Eq. \eqref{Eq:Ulambda}] with respect to $\tau$ yields the following equation of motion:
\begin{align}
\frac{d\hat{U}_{\Gu}(\tau,0)}{d \tau} =& \left(-\frac{i}{\hbar} \hat{H}(\tau)+\sum_{n=1}^\infty \frac{(i \Gu)^n}{n!} \hat{C}_{n}(\tau)\right)\hat{U}_{\Gu}(\tau,0),
\end{align}
where $\hat{C}_1(\tau) = \partial_\tau \hat{H}(\tau)$, $\hat{C}_2(\tau) = \left[ \hat{H}(\tau),\partial_\tau \hat{H}(\tau)\right]$, $\hat{C}_3(\tau) = \left[ \hat{H}(\tau),\left[ \hat{H}(\tau),\partial_\tau \hat{H}(\tau)\right]\right]$, etc. The generating function can be then written as (see Appendix \ref{A})
\begin{align}
G(\Gu) =&\text{Tr} \left\lbrace  \mathcal{T}_{\rightarrow} \exp \left({ \int_{0}^\tau dt   \sum_{n=1}^\infty (-1)^{n+1}\frac{(i \Gu)^n}{n!2^n} \hat{C}_{n}^{H}(t) } \right. \right) \nonumber \\ &\times \left. \mathcal{T}_{\leftarrow} \exp  \left({ \int_{0}^\tau dt \sum_{n=1}^\infty \frac{(i \Gu)^n}{n!2^n} \hat{C}_{n}^{H}(t)}  \right) \bar{\hat{\rho}}_0   \right\rbrace, \label{GeneC}
\end{align}
where the superscript $H$ indicates the Heisenberg picture, i.e., $\hat{C}_{n}^{H}(t)=\hat{U}^\dagger(t,0) \hat{C}_{n}(t) \hat{U}(t,0)$. The moments of work are then obtained by differentiating $G(\Gu)$ with respect to $\Gu$ at $\Gu=0$:
\begin{eqnarray}
\langle W^n \rangle = (-i)^n \left. \frac{\partial^n G(\Gu)}{\partial \Gu^n}\right|_{\Gu=0}. \label{Eq:moms}
\end{eqnarray}

With the implicit assumption that $[\hat{H}(t), \partial_t\hat{H}(t)]=0 $ (Ref. \onlinecite{Esposito2009}), 
$\hat{C}_n=0$ for $n > 1$, and the generating function becomes
\begin{eqnarray}
G_0(\Gu) =\text{Tr} \left\lbrace \mathcal{T}_{\rightarrow} \exp \left( { i\frac{\Gu}{2} \int_{0}^\tau dt \hat{P}^H(t)}\right)  \right. \nonumber \\   \times  \left. \mathcal{T}_{\leftarrow} \exp\left( { i\frac{\Gu}{2} \int_{0}^\tau dt \hat{P}^H(t)} \right) \bar{\hat{\rho}}_0   \right\rbrace, \label{Eq:Gene2}
\end{eqnarray}
where the power operator $\hat{P}$ (Ref. \onlinecite{Solinas2013}) is the time derivative of the total Hamiltonian, i.e.,  $\hat{P}^H(t)=\hat{U}^\dagger(t,0) \partial_t \hat{H}(t) \hat{U}(t,0)$.

The generating functions of Eqs. \eqref{GeneC} and \eqref{Eq:Gene2} are equivalent to the first order  of $\Gu$. Thus, both generating functions trivially 
give the same expression for the first moment of work as
\begin{align}
\langle W \rangle &= \int_0^\tau dt_1 \langle \hat{P}^H(t_1) \rangle. \label{Eq:4moms}
\end{align}
 Although the generating functions of Eqs. \eqref{GeneC} and \eqref{Eq:Gene2} differ already to second order in $\Gu$, 
the expressions for the second moment turn out to be equal as the corrections given by Eq. \eqref{GeneC} cancel out:
\begin{align}
\langle W^2 \rangle &= 2 \int_0^\tau dt_1 \int_0^{t_1} dt_2 \text{Re}\left\lbrace \langle \hat{P}^H(t_1) \hat{P}^H(t_2) \rangle \right\rbrace,
\end{align}
where we have used the Hermiticity of $\hat{P}$ to further simplify the expression. The same expressions for the first and second moment are also obtained by using the work operator with and without the commutator of the Hamiltonian at different times \cite{Engel2007}. However, for the third moment, the two generating functions give different results as
\begin{align}
\langle W^3 \rangle_0 &= 3 \mathclap{\int_{0}^{\tau}}dt_1 \mathclap{\int_{0}^{t_1}} dt_2 \mathclap{\int_{0}^{t_2}} dt_3 \text{Re}\left\lbrace \langle \hat{P}^H(t_1) \hat{P}^H(t_2) \hat{P}^H(t_3)\rangle \right. \nonumber \\  &+ \left. \langle \hat{P}^H(t_3) \hat{P}^H(t_1) \hat{P}^H(t_2)\rangle \right\rbrace, \\
\langle W^3 \rangle 
&= \langle W^3 \rangle_0+\frac{1}{4} \mathclap{\int_{0}^{\tau}}dt  \langle \hat{C}_{3}^{H}(t) \rangle \nonumber \\ &+\frac{3}{2}\int_{0}^{\tau}dt_1 \int_{0}^{t_1} dt_2 \text{Re}\left\lbrace  \langle  \hat{C}_{1}^{H}(t_1) \hat{C}_{2}^{H}(t_2) \rangle \right\rbrace,  
\end{align}
where $\langle W^3 \rangle_0$ denotes the third moment given by Eq. \eqref{Eq:Gene2} and $\langle W^3 \rangle$ denotes the one given 
by our general expression of Eq. \eqref{GeneC}.  The moments given by Eq. \eqref{Eq:Gene2} consist of third-order correlation functions of the power operator. 
In our result here, there are additional correction terms that involve commutators between the Hamiltonian and its time derivative, as expected. Such
correction terms appear also in the higher moments of work.

\section{Open quantum system}

To illustrate the importance of the results we have derived here, let us consider the special case of a weakly driven system, which is also weakly coupled to a heat bath \cite{Hekking2013}. 
The total Hamiltonian is taken to be of the form
\begin{equation}
\hat{H}(t)=\hat{H}_S(t)+\hat{H}_B+\hat{H}_{C},
\end{equation}
where subscripts $S, B$, and $C$ denote the system, bath, and bath-system 
interaction (coupling) Hamiltonians, respectively. Both the bath and the system-bath interaction (coupling) Hamiltonians are assumed to be time independent. The system Hamiltonian $\hat{H}_S(t)=\hat{H}_0+\hat{V}(t)$ consists of a time-independent part $\hat{H}_0$ and a time-dependent perturbative part  $\hat{V}(t)$. Therefore, the time derivative of the total Hamiltonian is simply 
given by $\hat{P}(t)=\partial_t \hat{H}(t) = \partial_t \hat{V}(t)$. 
In principle, we can calculate the moments of work from Eq. \eqref{GeneC}. However, already all the correlation functions of the third moment 
cannot be calculated just using the reduced density matrix $\hat{\rho}(t)$, as the correlation functions contain the total Hamiltonian 
that does not depend only on the system degrees of freedom but also on the bath degrees of freedom. To proceed, we consider a specific model, where a two-level system as in Ref. \onlinecite{Hekking2013} is bilinearly coupled to a heat bath of bosonic modes. The system Hamiltonian has the form 
\begin{eqnarray}
\hat{H}_S(t)&=&\hat{H}_0+\hat{V}(t); \\
\hat{H}_0&=&\hbar \omega_0 \hat{a}^\dagger \hat{a}; \\
\hat{V}(t)&=& \lambda(t) (\hat{a}^\dagger+\hat{a}),
\end{eqnarray}
where $a^\dagger =|e\rangle \langle g|$ and $a =|g\rangle \langle e|$ are the creation
and annihilation operators, respectively, in the ground-state ($|g\rangle$) and excited-state ($|e\rangle$) basis of
the undriven system, $\hbar \omega_0$ is the energy separation of the two
levels, and $\lambda(t)$ is the time-dependent drive. 
Further, the interaction and bath Hamiltonians are assumed to be of the form
\begin{eqnarray}
\hat{H}_{C}&=&\sum_k  (\hat{a}^\dagger +
\hat{a})  \otimes (g_k\hat{b}_k^\dagger +g^*_k
\hat{b}_k); \label{Eq:Hint} \\
\hat{H}_B&=&\sum_k \hbar \omega_k \hat{b}_k^\dagger \hat{b}_k,
\end{eqnarray}
where $g_k$ is the coupling strength, and $\hat{b}_k$ and $\hat{b}^\dagger_k$ are the bath annihilation and creation operators associated 
with energy $\hbar \omega_k$, respectively. For the total Hamiltonian $\hat{H}(t)$, this implies $[\hat{H}(t), \partial_t\hat{H}(t)] \neq 0$.  In the calculations, we approximate the initial density matrix $\bar{\hat{\rho}}_0$  with the tensor product of the system and bath density matrices, where both the system and the heat bath start in thermal equilibrium. That is, we neglect the interaction Hamiltonian in the energy measurements. Due to the weak driving and coupling to the heat bath, the evolution of the two level system can be 
approximated with the following Lindblad equation by invoking the Born-Markov and secular approximations (see Appendix \ref{B}):
\begin{eqnarray}
\frac{d \hat{\rho}}{dt}= &-&\frac{i}{\hbar} \left[ \hat{H}_S(t),\hat{\rho}(t)\right]  \nonumber \\ &+& \Gamma_{\downarrow} \left( {\rho}_{ee}(t)  | g\rangle \langle g |- \frac{1}{2} \left\lbrace \hat{\rho}(t), |e\rangle \langle e| \right\rbrace  \right) \nonumber \\ &+& \Gamma_{\uparrow} \left( {\rho}_{gg}(t)  |e\rangle \langle e|- \frac{1}{2} \left\lbrace \hat{\rho}(t), |g\rangle \langle g| \right\rbrace  \right), \label{Eq:ME}
\end{eqnarray}
where $\Gamma_{\downarrow}$ and $\Gamma_{\uparrow}=\Gamma_{\downarrow} e^{-\beta \hbar \omega_0}$ are the photon emission and absorption transition rates, respectively, $\hat{\rho}(t)$ is the density matrix of the reduced system in the Schr\"odinger picture and 
${\rho}_{kl}(t)= \langle k \vert \hat{\rho}(t) \vert l \rangle$.

As the secular approximation neglects the fast oscillating coupling terms, the same master equation could have been achieved by starting with the following form of the interaction Hamiltonian:
\begin{eqnarray}
 \hat{H}^{RWA}_{C}=\sum_k  g_k \hat{a}  \otimes \hat{b}_k^\dagger  +g^*_k
  \hat{a}^\dagger \otimes \hat{b}_k, \label{Eq:HRWA}
\end{eqnarray}
where the rotating wave approximation (RWA) has been invoked. With this form of the interaction Hamiltonian [Eq. \eqref{Eq:HRWA}], the jumps can be easily interpreted as photon emission and absorption to the bath. The usual quantum jump method \cite{Gardiner1992,Molmer1993,Carmichael1993, Plenio1998} can then be used to calculate the work distribution by interpreting the jumps as photon exchange while
neglecting the energetic contribution due to $\hat{H}_{C}$.

The first two moments for the system can be calculated in the usual manner by using the master equation of the reduced density matrix  as the operators in the correlation functions depend only on the system degrees of freedom \cite{Gardiner2004,Breuer2002}. For the third moment $\langle W^3 \rangle $, we can simplify the expression by using the fact that the power operator $\hat{P}(t)$ and the interaction Hamiltonian $\hat{H}_C$ [Eq. \eqref{Eq:Hint}] commute,
\begin{eqnarray}
\langle W^3 \rangle &=& \langle W^3 \rangle_0+ \frac{1}{4} \mathclap{\int_{0}^{\tau}}dt  \langle \left[  \hat{H}_{S}(t), \left[  \hat{H}_{S}(t), \hat{P}(t)\right] \right]  \rangle \nonumber \\  &+&\frac{3}{2}\int_{0}^{\tau}dt_1 \int_{0}^{t_1} dt_2 \text{Re}\left\lbrace  \langle \hat{P}(t_1) \left[  \hat{H}_{S}(t_2), \hat{P}(t_2)\right]  \rangle \right\rbrace \nonumber \\ &+&\frac{1}{4} \mathclap{\int_{0}^{\tau}}dt  \langle \left[  \hat{H}_{C}(t), \left[  \hat{H}_{S}(t), \hat{P}(t)\right] \right]  \rangle \nonumber \\
&\equiv& \langle W^3 \rangle_S+ \langle W^3 \rangle_{S+B} ,
\end{eqnarray}
where $\langle W^3 \rangle_S$ is given in the first two lines of the above equation and consists of the correlation functions that include only system operators. 
The interesting part is the second term $\langle W^3 \rangle_{S+B}$  that contains also operators that depend on the bath degrees of freedom,
\begin{eqnarray}
\langle W^3 \rangle_{S+B} = \frac{1}{4} \mathclap{\int_{0}^{\tau}}dt  \langle \left[  \hat{H}_{C}(t), \left[  \hat{H}_{S}(t), \hat{P}(t)\right] \right]  \rangle. \label{Eq:W3SB}
\end{eqnarray}
\begin{figure}[t]
                \centering
                \includegraphics[width=\columnwidth]{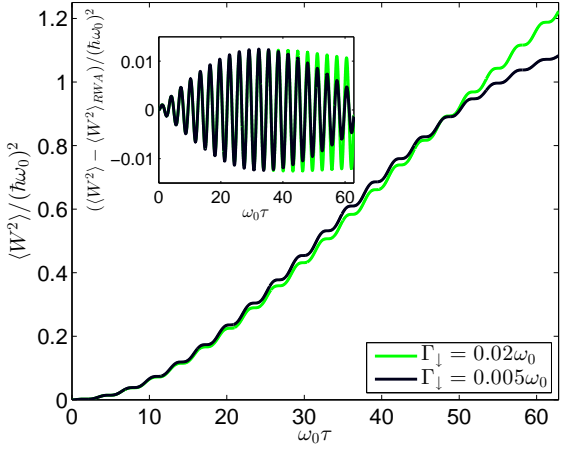}
        \caption{The numerical master equation results for the second moment 
        $\langle W^2 \rangle$ as a function of time for two different 
coupling strengths. Inset: The numerical results are compared to the analytical approximation
$\langle W^2 \rangle_{RWA}$ achieved with the additional RWA. The driving is assumed to be in resonance 
with $\omega_0$, i.e., $\omega=\omega_0$, $\beta \hbar \omega_0 = 2.0$, and $\lambda_0= 0.05 \hbar \omega_0$. The oscillation in the numerical results is 
caused by the fast oscillating terms of the drive. These are neglected in the analytical results by invoking the additional RWA. 
Inset: The oscillation for both coupling strengths is almost identical up to $ \omega_0 \tau =10 \pi $. \label{Fig:RWA}}
\end{figure}

We can estimate the term $\langle W^3 \rangle_{S+B}$ by invoking 
approximations similar to those used in the derivation of the corresponding master equation 
(see Appendix \ref{C}), yielding
\begin{eqnarray}
\langle W^3 \rangle_{S+B} &\approx&\frac{ \hbar^2 \omega_0}{2}  \left( \Gamma_{\uparrow}+\Gamma_{\downarrow}\right)  \int_0^\tau dt \dot{\lambda}(t) \text{Im}\left\lbrace {\rho}_{eg}(t) \right\rbrace. \label{Eq:SB}
\end{eqnarray}
Equation \eqref{Eq:SB} does not contain any bath degrees of freedom and can be calculated by solving the dynamics of the reduced system. 
With this form of $\langle W^3 \rangle_{S+B} $, the first three moments of work can be calculated numerically by 
using the master equation for a weak $\lambda(t)$. 

In the case of a simple sinusoidal resonance drive $\lambda(t)= \lambda_0 \sin (\omega_0 t)$, $\langle W^3 \rangle_{S+B}$ can be further 
approximated by changing to the interaction picture and neglecting the fast oscillating terms:
\begin{eqnarray}
 \langle W^3 \rangle_{S+B}  &\approx & \frac{\lambda_0 \hbar^2  \omega_0^2}{4 }   \left( \Gamma_{\uparrow}+\Gamma_{\downarrow}\right) \int_0^\tau dt  \text{Im}\left\lbrace {\rho}^I_{eg}(t)\right\rbrace, \label{Eq:SBRWA}
\end{eqnarray}
where $\hat{\rho}^I(t)$ is the density matrix of the reduced system in the interaction picture  with respect to $\hat{H}_0+\hat{H}_B$ and 
${\rho}^I_{eg}(t)= \langle e \vert \hat{\rho}^I(t) \vert g \rangle$.

\begin{figure}[t]
                \centering
                \includegraphics[width=\columnwidth]{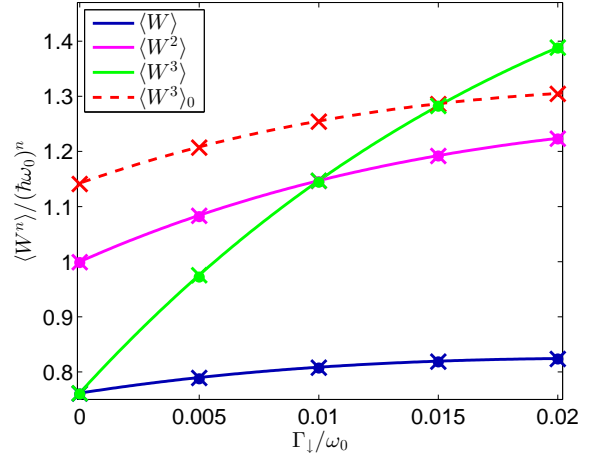}
        \caption{Comparison of the quantum jump method and master-equation results for the first three moments for different coupling amplitudes. 
The solid and dashed lines correspond to the analytical results with the additional RWA, the dots correspond to the numerical quantum jump results, 
and the crosses correspond to the numerical master-equation data.  The driving is assumed to be in resonance with 
$\omega_0$, i.e., $\omega=\omega_0$, $\beta \hbar \omega_0 = 2.0$, $\lambda_0= 0.05 \hbar \omega_0$, and the drive lasts for $10$ cycles, i.e., $\omega \tau = 20 \pi$.  
The numerical results are calculated with $10^4$ time steps. The quantum jump results consist of $10^6$ realizations. The numerical master-equation and quantum jump results give a good agreement within the numerical accuracy: The largest difference in $\langle W^n \rangle/(\hbar \omega_0)^n$ is less than $0.0032$.   \label{Fig:CMP}}
\end{figure}

For the sinusoidal resonance drive, we can simplify the analytical calculations of the correlation functions of the work moments with 
an additional rotating wave approximation. By neglecting the fast oscillating terms,  the power operator simplifies to the form 
$\hat{P}^I(t) \approx  \lambda_0 \omega_0 (\hat{a}+\hat{a}^\dagger )/2 $ in the interaction picture. Using the regression theorem 
\cite{Lax1963}, we can then calculate analytical approximations for the moments of work.

The regression theorem results with the additional RWA were found to give an excellent agreement with 
the numerical master equation results when the driving period $\tau$ consists of full or half cycles. When the driving period is not 
$\omega_0 \tau =n \pi$, where $n$ is an integer, then there can be a small difference between the regression theorem results and the numerical master equation results. This difference is due to the oscillation caused by the fast oscillation terms of the drive for the latter and is illustrated in Fig. \ref{Fig:RWA} for the second moment $\langle W^2 \rangle$ with $\lambda_0=0.05 \hbar \omega_0$.  
As the oscillation is caused by the fast oscillating terms of the drive, the deviation becomes larger when the value of $\lambda_0$ is increased.

We also compared the values of the first three moments of 
$G(\Gu)$ [Eq. \eqref{GeneC}] to the quantum jump results. 
Our results and the quantum jump method results are in good agreement within the numerical accuracy 
for all of the first three moments independently of the 
parametric values, as illustrated in Fig. \ref{Fig:CMP}.
We also calculated and found our results to be in agreement with the generalized
master-equation results \cite{Esposito2009,Silaev2013}. The results are also in accordance with the ones of Ref. \onlinecite{Liu2014}.

\begin{figure}[t]
                \centering
                \includegraphics[width=\columnwidth]{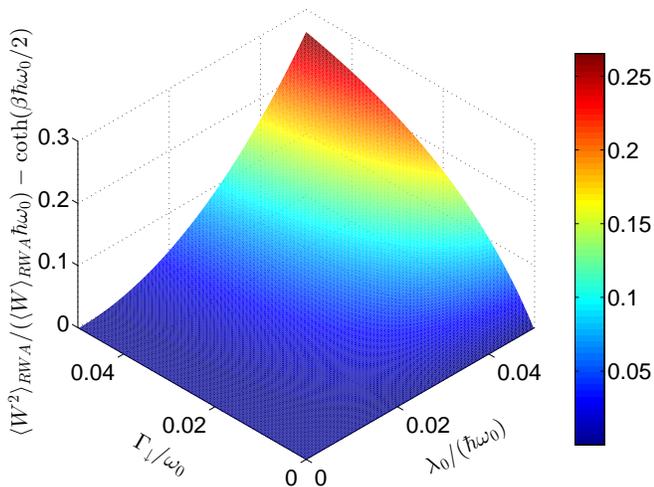}
        \caption{Test of the standard fluctuation dissipation theorem  ($\langle W^2 \rangle_{\text{\tiny{RWA}}}/\langle W \rangle_{\text{\tiny{RWA}}}=\hbar \omega_0 \coth (\beta \hbar \omega_0/2)$ for different coupling and driving amplitudes. Here, the driving is assumed to be in resonance 
        with $\omega_0$, i.e., $\omega=\omega_0$, $\beta \hbar \omega_0 = 2.0$, and the drive lasts for $10$ cycles, i.e., $\omega \tau = 20 \pi$.  
        As expected, significant deviations start appearing with increased coupling and drive. \label{fig:W2a}}
\end{figure}

The third moments of both generating functions (Eqs. \eqref{GeneC} and \eqref{Eq:Gene2}) are 
presented in Fig. \ref{Fig:CMP} as well. Clearly, the third moment without the 
correction, $\langle W^3 \rangle_0$, differs greatly from $\langle W^3 \rangle$ and the quantum jump results even in the case of no coupling to the heat bath.  
From the correction terms, the term $\langle W^3 \rangle_{S+B}$ [Eq. \eqref{Eq:SB}] was found to be several orders of magnitude smaller
than $\langle W^3 \rangle_0$ for the weakly driven system here. 
In the regression theorem results with the additional RWA, $\langle W^3 \rangle_{S+B}$ [Eq. \eqref{Eq:SBRWA}] 
is always zero, as the density matrix remains real in the interaction picture.

In Fig. \ref{fig:W2a}, we further illustrate the expected deviation from the standard fluctuation dissipation theorem\cite{Callen1951}  
(FDT) $\langle W^2 \rangle_{RWA} / \langle W \rangle_{RWA} = \hbar \omega_0 \coth (\beta \hbar \omega_0/2)$ for large drive 
amplitudes and coupling strengths. 
From Fig. \ref{fig:W2a}, we see that the FDT is valid  not only in the linear response regime ($\lambda_0 \rightarrow 0$) 
but also in the limit of no coupling ($\Gamma_\downarrow \rightarrow 0$)  with arbitrary drive amplitude within this model. In the case of no coupling, the probability to end up in the excited state when starting from the  ground state, denoted as $p_{ge}=|\langle e | \hat{U}(\tau,0) |g\rangle |^2$,  is exactly the same as the probability to end up in the ground state when starting from the excited state, $p_{eg}$. Hence, $\langle W^n \rangle = (\hbar \omega_0)^n \rho_{gg}(0) p_{ge} +(-\hbar \omega_0)^n \rho_{ee}(0) p_{eg}  =  (\hbar \omega_0)^n p_{ge}  (\rho_{gg}(0)+(-1)^n \rho_{ee}(0))$, which immediately gives the FDT when we start from thermal equilibrium.

For small values of 
$\lambda_0$ and $\Gamma_\downarrow$, the deviation from the FDT increases almost parabolically when the drive 
amplitude $\lambda_0$ increases and the transition rate $\Gamma_\downarrow$ remains 
constant for small values of $\lambda_0$ and $\Gamma_\downarrow$. 
This can be seen by Taylor expanding  $\langle W^2 \rangle_{RWA} / \langle W \rangle_{RWA} $ 
around $(\lambda_0,\Gamma_\downarrow)=(0,0)$,
\begin{eqnarray}
\langle W^2 \rangle_{RWA} / \langle W \rangle_{RWA} = \hbar \omega_0 \coth (\beta \hbar \omega_0/2) \nonumber \\
+ \hbar \omega_0 \Gamma_\downarrow \frac{\lambda_0^2 \tau^3}{60 \hbar^2} (1-e^{-\beta \hbar \omega_0}) (1-\Gamma_\downarrow \frac{\tau}{6}(1+e^{-\beta \hbar \omega})) \nonumber \\+\mathcal{O}((\Gamma_\downarrow \tau)^3)+\mathcal{O}((\lambda_0 \tau / \hbar )^4). \label{approx}
\end{eqnarray}
This expansion is valid up to  $\Gamma_\downarrow , \lambda_0 / \hbar \lesssim 0.01 \omega_0$ in Fig. \ref{fig:W2a} 
as the higher-order terms become important already when $\Gamma_\downarrow , \lambda_0 / \hbar  = 0.01 \omega_0$, 
due to the high number of drive cycles.

\section{Summary and Conclusions}

In summary, we have examined in detail the distribution of work done when a two-measurement protocol is applied to a driven
open quantum system. To this end, we have first derived a general form for the generating function of work
and studied the first three moments of work by using the master equation of the reduced system and 
invoking approximations similar to the ones made in the microscopic derivation of the reduced density matrix. 
We have compared our results to the earlier derivations \cite{Esposito2009} that were carried out implicitly assuming that the total Hamiltonian 
and its time derivative commute and have shown that there is a significant difference already in the case of the third moment. 
This emphasizes the importance of properly evaluating the higher moments of work, which are needed to check fluctuation relations
such as the JE. To make our results concrete, we have considered a weakly driven and weakly coupled two-level
system by using a number of different techniques, including the quantum jump method. Our results demonstrate the influence of the correct choice of the generating function already in the results for the third moment of work distribution.

\section{Acknowledgements}
We thank F. Hekking, T. Sagawa, I. Savenko and A. Kutvonen for discussions. This work has been supported in part by the V\"ais\"al\"a Foundation, the 
European Union Seventh Framework Programme INFERNOS (FP7/2007-2013) under Grant No.
308850, and the Academy of Finland through its Centre of Excellence Programs 
(Projects No. 250280 and No. 251748). P.S. acknowledges financial support from 
FIRB-Futuro in Ricerca 2013 under Grant No. RBFR1379UX and FIRBFuturo in Ricerca 2012 under Grant No. 
RBFR1236VV HybridNanoDev.

\appendix

\section{Generating function of the two-point measurement protocol} \label{A}

In the two-measurement protocol for a closed quantum system, the probability to measure $E_0$ at time $t=0$ and $E_{\tau}$ at $t=\tau$ is of the form
\begin{equation}
P[E_\tau,E_0]=\text{Tr}\lbrace \hat{P}_{E_\tau} \hat{U}(\tau,0) \hat{P}_{E_0} \hat{\rho}_0 \hat{P}_{E_0} \hat{U}^\dagger(\tau,0) \hat{P}_{E_\tau}  \rbrace, \label{AEq:prob}
\end{equation}
where $\hat{U}(\tau,0)=\mathcal{T}_\leftarrow \exp{\left(-\frac{i}{\hbar} \int_0^\tau dt \hat{H}(t) \right) }$ is the unitary time evolution operator, $\mathcal{T}_\leftarrow$ describes the chronological time ordering, and the projection operators are given by $\hat{P}_{E_t}=| E_t \rangle \langle E_t |$, where $| E_t \rangle$ is the state corresponding to the measurement result $E_{t}$ at time $t$. The corresponding generating function is given by \cite{Esposito2009}
\begin{align}
&G(\Gu)=\sum\limits_{E_0,E_\tau} e^{i \Gu (E_\tau-E_0)} P[E_\tau,E_0] \\
	 &=\sum\limits_{E_\tau} \text{Tr}\left\lbrace \hat{U}(\tau,0) \sum\limits_{E_0} \left(  e^{-i (\Gu /2) E_0} \hat{P}_{E_0}  \hat{\rho}_0 \hat{P}_{E_0} e^{-i (\Gu /2) E_0} \right)  \right. \nonumber \\   &\times \left. \hat{U}^\dagger(\tau,0) \hat{P}_{E_\tau} e^{i \Gu  E_\tau} \right\rbrace \nonumber \\
	&= \text{Tr} \left\lbrace \hat{U}(\tau,0)  e^{-i (\Gu /2) \hat{H}(0)}  \bar{\hat{\rho}}_0 e^{-i (\Gu /2) \hat{H}(0)} \hat{U}^\dagger(\tau,0) e^{i \Gu \hat{H}(\tau)} \right\rbrace \nonumber \\ 
	 &= \text{Tr} \left\lbrace \hat{U}_{\Gu /2}(\tau,0)  \bar{\hat{\rho}}_0 \hat{U}_{-\Gu/2}^\dagger(\tau,0)  \right\rbrace,  \label{AEq:Gene}
\end{align}
where
\begin{eqnarray}
\hat{U}_{\Gu}(\tau,0)&=&e^{i \Gu \hat{H}(\tau)} \hat{U}(\tau,0)  e^{-i \Gu \hat{H}(0)}; \label{AEq:Ulambda} \\ \bar{\hat{\rho}}_0&=&\sum_{E_0} \hat{P}_{E_0}\hat{\rho}_0 \hat{P}_{E_0},
\end{eqnarray}
and $\hat{\rho}_0$ is the initial density matrix. If the initial density matrix is diagonal in the first measurement's basis, 
then $\bar{\hat{\rho}}_0=\hat{\rho}_0$. In the case of energy measurement 
this means that if the total system Hamiltonian $\hat{H}(0)$ commutes with $\hat{\rho}_0$, e.g., 
the density matrix is diagonal in the eigenbasis of $\hat{H}(0)$, then $\bar{\hat{\rho}}_0=\hat{\rho}_0$.

With the assumption $[\hat{H}(t), \partial_t\hat{H}(t)]=0 $, the evolution operator $\hat{U}_{\Gu}(\tau,0)$ satisfies the following equation of motion:
\begin{equation}
\frac{d}{d \tau} \hat{U}_{\Gu}(\tau,0) = - \frac{i}{\hbar} (\hat{H}(\tau)- \hbar  \Gu \frac{\partial \hat{H}(\tau)}{\partial \tau}) \hat{U}_{\Gu}(\tau,0). \label{AEq:dU}
\end{equation}
Since $\hat{U}_{\Gu}(0,0)=\hat{1}$, $\hat{U}_{\Gu}(\tau,0)$ can be expressed as
\begin{equation}
\hat{U}_{\Gu}(\tau,0) = \mathcal{T}_{\leftarrow} \exp \left[{- \frac{i}{\hbar} \int_{0}^\tau dt \left(\hat{H}(t)-\hbar\Gu \frac{\partial}{\partial t} \hat{H}(t) \right)}\right] . \label{AEq:Ul}
\end{equation}
However, contrary to Ref. \onlinecite{Esposito2009}, this solution\cite{footnote2} is not the general one due to the implicit assumption that 
$[\hat{H}(t), \partial_t\hat{H}(t)]=0 $. With this form of $\hat{U}_{\Gu/2}(\tau,0)$, the generating function simplifies to
\begin{align}
G_0(\Gu) &=\text{Tr} \left\lbrace \mathcal{T}_{\leftarrow} \exp \left[ {- \frac{i}{\hbar} \int_{0}^\tau dt \left( \hat{H}(t)-\hbar \frac{\Gu}{2} \frac{\partial}{\partial t} \hat{H}(t) \right)} \right] \bar{\hat{\rho}}_0 \right.  \nonumber \\ &\times \left.\mathcal{T}_{\rightarrow} \exp \left[{ \frac{i}{\hbar} \int_{0}^\tau dt \left(\hat{H}(t)+\hbar \frac{\Gu}{2}\frac{\partial}{\partial t} \hat{H}(t)\right)} \right] \right\rbrace . \label{AEq:Gene1a}
\end{align}
Let us denote the time derivative of the total Hamiltonian as the power operator 
$\hat{P}(t)=\partial \hat{H}(t)/ \partial t $. In order to get an expression where the operators are expressed in the Heisenberg picture, we can use the unitarity of $\hat{U}(\tau,0)$ and calculate the equation of motion for the operators 
$\hat{U}^\dagger(\tau,0)\hat{U}_{\Gu/2}(\tau,0)$ and $\hat{U}^\dagger_{-\Gu/2}(\tau,0)\hat{U}(\tau,0)$.  
Changing to this Heisenberg picture and using the periodicity of the trace then gives the final form,
\begin{align}
G_0(\Gu) =&\text{Tr} \left\lbrace \mathcal{T}_{\rightarrow} \exp \left( { i\frac{\Gu}{2} \int_{0}^\tau dt \hat{P}^H(t)}\right)  \right. \nonumber \\ &\times \left. \mathcal{T}_{\leftarrow} \exp\left( { i\frac{\Gu}{2} \int_{0}^\tau dt \hat{P}^H(t)} \right) \bar{\hat{\rho}}_0   \right\rbrace, \label{AEq:Gene2}
\end{align}
where $\hat{P}^H(t)=\hat{U}^\dagger(t,0) (\partial \hat{H}(t)/ \partial t)  \hat{U}(t,0)$. 

Without the assumption $[\hat{H}(t), \partial_t\hat{H}(t)]=0 $, the differentiation of the evolution operator $\hat{U}_{\Gu}(\tau,0)$ [Eq. \eqref{AEq:Ulambda}] with respect to $\tau$ yields 
\begin{align}
\frac{d \hat{U}_{\Gu}(\tau,0)}{d \tau} =& \left(-\frac{i}{\hbar} \hat{H}(\tau)+\sum_{n=1}^\infty \frac{(i \Gu)^n}{n!} \hat{C}_{n}(\tau)\right)\hat{U}_{\Gu}(\tau,0),
\end{align}
where $\hat{C}_1(\tau) = \partial_\tau \hat{H}(\tau)$, $\hat{C}_2(\tau) = \left[ \hat{H}(\tau),\partial_\tau \hat{H}(\tau)\right]$, $\hat{C}_3(\tau) = \left[ \hat{H}(\tau),\left[ \hat{H}(\tau),\partial_\tau \hat{H}(\tau)\right]\right]$, etc. Similarly,
\begin{align}
\frac{d \hat{U}^\dagger_{\Gu}(\tau,0)}{d \tau}
&= \hat{U}^\dagger_{\Gu}(\tau,0) \left( \frac{i}{\hbar} \hat{H}^\dagger(\tau)+ \sum_{n=1}^\infty \frac{(-i \Gu)^n}{n!} \hat{C}^\dagger_{n}(\tau)\right) \\
&= \hat{U}^\dagger_{\Gu}(\tau,0) \left( \frac{i}{\hbar} \hat{H}(\tau)- \sum_{n=1}^\infty \frac{(i \Gu)^n}{n!} \hat{C}_{n}(\tau)\right).
\end{align}
Again, since $\hat{U}_{\Gu}(0,0)=\hat{U}^\dagger_{\Gu}(0,0)=\hat{1}$, the operators $\hat{U}_{\Gu}(\tau,0)$ and $\hat{U}^\dagger_{\Gu}(\tau,0)$ can be expressed as follows:
\begin{align}
\hat{U}_{\Gu}(\tau,0) &= \mathcal{T}_{\leftarrow} \exp \left[{\int_{0}^\tau dt \left(-\frac{i}{\hbar} \hat{H}(t)+\sum_{n=1}^\infty \frac{(i \Gu)^n}{n!} \hat{C}_{n}(t)\right)}\right]; \label{AEq:Ul2} \\
\hat{U}^\dagger_{\Gu}(\tau,0) &= \mathcal{T}_{\rightarrow} \exp \left[ {\int_{0}^\tau dt \left( \frac{i}{\hbar} \hat{H}(t)- \sum_{n=1}^\infty \frac{(i \Gu)^n}{n!} \hat{C}_{n}(t)\right) } \right].
\end{align}
After changing to the Heisenberg picture described earlier, the exact generating function reads
\begin{align}
G(\Gu) =&\text{Tr} \left\lbrace  \mathcal{T}_{\rightarrow} \exp \left({ \int_{0}^\tau dt   \sum_{n=1}^\infty (-1)^{n+1}\frac{(i \Gu)^n}{n!2^n} \hat{C}_{n}^{H}(t) } \right. \right) \nonumber \\ &\times \left. \mathcal{T}_{\leftarrow} \exp  \left({ \int_{0}^\tau dt \sum_{n=1}^\infty \frac{(i \Gu)^n}{n!2^n} \hat{C}_{n}^{H}(t)}  \right) \bar{\hat{\rho}}_0   \right\rbrace, \label{AGeneC}
\end{align}
where $\hat{C}_{n}^{H}(t)=\hat{U}^\dagger(t,0) \hat{C}_{n}(t) \hat{U}(t,0)$.

\section{Calculation of the master equation} \label{B}

Let us denote the density matrix of the total system with $\hat{\rho}_{T}(t)$. 
The density matrix of the reduced system  $\hat{\rho}(t)$  is obtained by tracing over the bath degrees of freedom,
\begin{eqnarray}
\hat{\rho}(t)&=&\text{Tr}_B\left\lbrace \hat{\rho}_{T}(t) \right\rbrace.
\end{eqnarray}

Similarly, the density matrix of the bath  $\hat{\rho}_B(t)$  is obtained by tracing over the system degrees of freedom,
\begin{eqnarray}
\hat{\rho}_B(t)&=&\text{Tr}_S\left\lbrace \hat{\rho}_{T}(t) \right\rbrace.
\end{eqnarray}
The Hamiltonian of the total closed system can be written as
\begin{equation}
\hat{H}(t)=\hat{H}_0+\hat{H}_B+\hat{V}(t)+\hat{H}_{C}.
\end{equation}
Let us change to the interaction picture with respect to $(\hat{H}_0+\hat{H}_B)$, denoted by the superscript $I$. We can write the equation of motion for the total density matrix as
\begin{align}
\frac{ d\hat{\rho}^I_{T}(t)}{dt}&= -\frac{i}{\hbar}  \left[ \hat{V}^I(t), \hat{\rho}^I_{T}(t) \right] -\frac{i}{\hbar}  \left[ \hat{H}^I_C(t), \hat{\rho}^I_{T}(0) \right]  \nonumber \\ &-\frac{1}{\hbar^2} \int_0^t dt^\prime \left[ \hat{H}^I_{C}(t), \left[ \hat{V}^I(t^{\prime})+ \hat{H}^I_{C}(t^\prime), \hat{\rho}^I_{T}(t^\prime) \right]  \right]. \label{Eq:MET0}
\end{align}

We will approximate the initial density matrix after the first measurement with $\hat{\rho}^I_T(0)=\hat{\rho}^I(0) \otimes \hat{\rho}^I_{B}(0)$, where both the system and the heat bath start in thermal equilibrium. This approximation corresponds to that of neglecting the interaction Hamiltonian in the energy measurements. A similar approximation is done also in the calculation of the moments. Tracing over the bath degrees of freedom, we get the following equation for the reduced density matrix:
\begin{align}
&\frac{ d\hat{\rho}^I(t)}{dt}  = -\frac{i}{\hbar}  \left[ \hat{V}^I(t), \hat{\rho}^I (t) \right]  \nonumber \\ &-\frac{i}{\hbar}  \text{Tr}_{B} \left\lbrace \left[ \hat{H}^I_C(t), \hat{\rho}^I(0) \otimes \hat{\rho}^I_{B}(0) \right] \right\rbrace  \nonumber \\ &-\frac{1}{\hbar^2} \int_0^t dt^\prime \text{Tr}_{B} \left\lbrace \left[ \hat{H}^I_{C}(t), \left[ \hat{V}^I(t^{\prime})+ \hat{H}^I_{C}(t^\prime), \hat{\rho}^I_{T}(t^\prime) \right]  \right] \right\rbrace. \label{Eq:MET}
\end{align}

Let us denote the last term on the right hand side of Eq. \eqref{Eq:MET} as $\chi(t)$. Invoking the Born approximation  [$\hat{\rho}^I_T(t)=\hat{\rho}^I(t) \otimes \hat{\rho}^I_{B}(0)$] and the Markov approximation, it changes to
 \begin{align}
\chi (t) &=  -\frac{1}{\hbar^2} \int_0^\infty dt^\prime \text{Tr}_{B} \left\lbrace \left[  \hat{H}^I_{C}(t), \left[  \hat{V}^I(t^\prime), \hat{\rho}^I(t) \otimes \hat{\rho}^I_{B}(0)  \right]  \right]  \right.\nonumber \\ &+ \left. \left[  \hat{H}^I_{C}(t), \left[ \hat{H}^I_{C}(t^\prime), \hat{\rho}^I(t) \otimes \hat{\rho}^I_{B}(0)  \right]  \right]  \right\rbrace .
\end{align}
 
The interaction Hamiltonian can be expressed as $\hat{H}^I_{C}(t)=\sum_j \hat{A}^I_j(t) \otimes \hat{B}^I_j(t) $, where $\hat{A}^I_j(t)$ acts on the system degrees of freedom and  $\hat{B}^I_j(t)$ acts on the bath degrees of freedom. With this expression of $\hat{H}^I_{C}(t)$ and assuming that $\text{Tr}_B \left\lbrace  \hat{B}^I_j(t)  \hat{\rho}_B^I(0) \right\rbrace = 0$, $\chi$ changes to the form%
\begin{eqnarray}
\chi(t) & = & -\frac{1}{\hbar^2} \int_0^\infty dt^\prime  \sum_{j,k}\left( \hat{A}^I_k(t) \hat{A}^I_j(t^\prime) \hat{\rho}^I(t) \right.  - \left. \hat{A}^I_j(t^\prime) \hat{\rho}^I(t) \hat{A}^I_k(t)  \right) \nonumber \\ &\times& \text{Tr}_B \left\lbrace  \hat{B}^I_k(t)\hat{B}^I_j(t^\prime) \hat{\rho}_B^I(0) \right\rbrace \nonumber \\   &+& \left(   \hat{\rho}^I(t) \hat{A}^I_j(t^\prime) \hat{A}^I_k(t) - \hat{A}^I_k(t) \hat{\rho}^I(t) \hat{A}^I_j(t^\prime) \right)  \nonumber \\ &\times&  \text{Tr}_B \left\lbrace  \hat{B}^I_j(t^\prime) \hat{B}^I_k(t) \hat{\rho}_B^I(0) \right\rbrace.  \label{AEq:C32}
\end{eqnarray}
For the system studied, the bath correlation functions are given by
\begin{eqnarray}
\text{Tr}_B \left\lbrace  \hat{B}^I(t)\hat{B}^I(t^\prime) \hat{\rho}_B^I(0) \right\rbrace &&= \sum_k |g_k|^2 \left[ e^{i \omega_k (t-t^\prime)} n_k \right. \nonumber \\ +&& \left. e^{-i \omega_k(t-t^\prime)}( n_k+1) \right],
\end{eqnarray}
where $\hat{B}^I(t)= \sum_k e^{-i \omega_k t} g_k \hat{b}+e^{i \omega_k t}  g_k^* \hat{b}^\dagger $ and $n_k$ is the average number of photons with frequency $\omega_k$. The expression of $\chi$ can be simplified by taking into account that $\int_0^\infty dt e^{i \omega t} = \pi \delta(\omega) +i \mathcal{P}(\frac{1}{\omega})$, where $\mathcal{P}$ denotes the Cauchy principal value and the imaginary part only affects the Lamb shift. By neglecting the Lamb shift and invoking the secular approximation, i.e., neglecting  the fast oscillating terms, we get
\begin{eqnarray}
\chi(t) &=&  \Gamma_{\downarrow} \left( {\rho}^I_{ee}(t)  | g\rangle \langle g |- \frac{1}{2} \left\lbrace \hat{\rho}^I(t), |e\rangle \langle e| \right\rbrace  \right) \nonumber \\ &+& \Gamma_{\uparrow} \left( {\rho}_{gg}^I(t)  |e\rangle \langle e|- \frac{1}{2} \left\lbrace \hat{\rho}^I(t), |g\rangle \langle g| \right\rbrace  \right), \label{Eq:MEC}
\end{eqnarray}
where $\hat{\rho}^I_{kl}(t)= \langle k \vert \hat{\rho}^I(t) \vert l \rangle$ and the transition rates are given by
\begin{eqnarray}
\Gamma_\downarrow&=& \frac{2 \pi}{ \hbar^2} \sum_k (n_k+1) |g_k|^2 \delta (\omega_0-\omega_k), \\
\Gamma_\uparrow&=& \frac{2 \pi}{ \hbar^2} \sum_k n_k |g_k|^2 \delta (\omega_0-\omega_k),
\end{eqnarray}
and they satisfy the detailed balance $\Gamma_\uparrow=\Gamma_\downarrow e^{-\beta \hbar \omega_0}$. With the approximation $\hat{\rho}^I_T(0)=\hat{\rho}^I(0) \otimes \hat{\rho}^I_{B}(0)$, the second term on the right-hand side of Eq. \eqref{Eq:MET} goes to zero due to $\text{Tr}_B \left\lbrace  \hat{B}^I(t)  \hat{\rho}_B^I(0) \right\rbrace = 0$. Thus, switching back to the Schr\"odinger picture gives us the following master equation: 
\begin{eqnarray}
\frac{d \hat{\rho}}{dt}= &-&\frac{i}{\hbar} \left[ \hat{H}_S(t),\hat{\rho}(t)\right]  \nonumber \\ &+& \Gamma_{\downarrow} \left( {\rho}_{ee}(t)  | g\rangle \langle g |- \frac{1}{2} \left\lbrace \hat{\rho}(t), |e\rangle \langle e| \right\rbrace  \right) \nonumber \\ &+& \Gamma_{\uparrow} \left( {\rho}_{gg}(t)  |e\rangle \langle e|- \frac{1}{2} \left\lbrace \hat{\rho}(t), |g\rangle \langle g| \right\rbrace  \right). \label{Eq:ME}
\end{eqnarray}

\section{Calculation of $\langle W^3 \rangle_{S+B}$} \label{C}

Using the same notation as in the derivation of the master equation, we can write the total density matrix in the interaction picture with respect to $(\hat{H}_0+\hat{H}_B)$ as
\begin{eqnarray}
\hat{\rho}^I_{T}(t)&=&\hat{\rho}^I_{T}(0) -\frac{i}{\hbar} \int_0^t dt^\prime \left[  \hat{H}^I_{C}(t^\prime)+\hat{V}^I(t^\prime), \hat{\rho}^I_{T}(t^\prime) \right]. \nonumber \\ 
\end{eqnarray}
With this form of $\hat{\rho}^I_{T}(t)$,  the term inside the integral in Eq. \eqref{Eq:W3SB} can be written as
\begin{eqnarray}
\Xi (t)  &\equiv& \langle \left[  \hat{H}^I_{C}(t), \left[  \hat{H}^I_{S}(t), \hat{P}^I(t)\right] \right]  \rangle \\
& = & \text{Tr}_{S+B}\left\lbrace \left[  \hat{H}^I_{C}(t), \left[  \hat{H}^I_{S}(t), \hat{P}^I(t)\right] \right] \hat{\rho}^I_T(t) \right\rbrace  \\
& = & \text{Tr}_{S+B}\left\lbrace \left[  \hat{H}^I_{C}(t), \left[  \hat{H}^I_{S}(t), \hat{P}^I(t)\right] \right] \hat{\rho}^I_T(0) \right\rbrace \nonumber \\
& - & \frac{i}{\hbar} \int_0^t dt^\prime \text{Tr}_{S+B}\left\lbrace \left[  \hat{H}^I_{C}(t), \left[  \hat{H}^I_{S}(t), \hat{P}^I(t)\right] \right] \right.  \nonumber \\ &\times&  \left. \left[  \hat{H}^I_{C}(t^\prime)+\hat{V}^I(t^\prime), \hat{\rho}^I_T(t^\prime)\right] \right\rbrace. 
\end{eqnarray}
Again, we will approximate the initial density matrix with $\hat{\rho}^I_T(0)=\hat{\rho}^I(0) \otimes \hat{\rho}^I_{B}(0)$, where both the system and the heat bath start in thermal equilibrium. Using the Born approximation [$\hat{\rho}^I_T(t)=\hat{\rho}^I(t) \otimes \hat{\rho}^I_{B}(0)$], we can approximate $\Xi (t)$ with%
\begin{eqnarray}
\Xi (t) & = & \text{Tr}_{S+B}\left\lbrace \left[  \hat{H}^I_{C}(t), \left[  \hat{H}^I_{S}(t), \hat{P}^I(t)\right] \right] \hat{\rho}^I(0) \otimes \hat{\rho}^I_{B}(0) \right\rbrace \nonumber \\
 & - &\frac{i}{\hbar} \int_0^t dt^\prime \text{Tr}_{S+B}\left\lbrace \left[  \hat{H}^I_{C}(t), \left[  \hat{H}^I_{S}(t), \hat{P}^I(t)\right] \right] \right.  \nonumber \\ &\times&  \left. \left[  \hat{H}^I_{C}(t^\prime)+\hat{V}^I(t^\prime), \hat{\rho}^I(t^\prime) \otimes \hat{\rho}^I_{B}(0) \right] \right\rbrace. 
\end{eqnarray}
The interaction Hamiltonian can be written as $\hat{H}_{C}^I(t)=\sum_j \hat{A}^I_j(t) \otimes \hat{B}^I_j(t) $, where $\hat{A}^I_j(t)$ acts on the system degrees of freedom and  $\hat{B}^I_j(t)$ acts on the bath degrees of freedom. Assuming $\text{Tr}_B \left\lbrace  \hat{B}^I_j(t)  \hat{\rho}_B^I(0) \right\rbrace = 0$, $\Xi (t)$ reduces to
\begin{eqnarray}
\Xi (t) & = &- \frac{i}{\hbar} \int_0^t dt^\prime \text{Tr}_{S+B}\left\lbrace \left[  \hat{H}^I_{C}(t), \left[  \hat{H}^I_{S}(t), \hat{P}^I(t)\right] \right] \right.  \nonumber \\ &\times&  \left. \left[  \hat{H}^I_{C}(t^\prime), \hat{\rho}^I(t^\prime) \otimes \hat{\rho}^I_{B}(0) \right] \right\rbrace. 
\end{eqnarray}

Invoking the Markov approximation, the expression changes to
\begin{eqnarray}
\Xi (t)  &=&-\frac{i}{\hbar} \int_0^\infty dt^\prime \text{Tr}_{S+B}\left\lbrace \left[  \hat{H}^I_{C}(t), \left[  \hat{H}^I_{S}(t), \hat{P}^I(t)\right] \right] \right.  \nonumber \\ &\times&  \left. \left[  \hat{H}^I_{C}(t^\prime), \hat{\rho}^I(t) \otimes \hat{\rho}_{B}^I(0) \right] \right\rbrace. \label{AEq:C31}
\end{eqnarray}
Expressing the interaction Hamiltonian as $\hat{H}_{C}^I(t)=\sum_j \hat{A}^I_j(t) \otimes \hat{B}^I_j(t) $ and denoting $\hat{Q}_j^I(t)= \left[\hat{A}^I_j(t), \left[  \hat{H}^I_{S}(t), \hat{P}^I(t) \right]  \right]$, Eq. \eqref{AEq:C31} changes to the form
\begin{eqnarray}
\Xi (t) &=&-\frac{i}{\hbar} \sum_{j,k} \int_0^\infty dt^\prime \left(  \text{Tr}_{S}\left\lbrace \hat{Q}^I_k(t)\hat{A}^I_j(t^\prime) \hat{\rho}^I(t) \right\rbrace \right.
\nonumber \\ &\times& \left. \text{Tr}_B \left\lbrace  \hat{B}^I_k(t)\hat{B}^I_j(t^\prime) \hat{\rho}_B^I(0) \right\rbrace \right. \nonumber \\ &-& \left. \text{Tr}_{S}\left\lbrace \hat{A}^I_j(t^\prime) \hat{Q}^I_k(t) \hat{\rho}^I(t) \right\rbrace \right. \nonumber \\ &\times& \left. \text{Tr}_B \left\lbrace  \hat{B}^I_j(t^\prime)\hat{B}^I_k(t) \hat{\rho}_B^I(0) \right\rbrace \right). \label{AEq:C32}
\end{eqnarray}
For the system studied, $\Xi(t)$ reduces to
\begin{eqnarray}
\Xi (t)  &=&-i2\omega_0 \dot{\lambda}(t) \left(  \rho^I_{eg}(t) \int_0^\infty dt^\prime e^{-i \omega_0 t^\prime} \xi(t,t^\prime) \right. \nonumber \\ &-&  \left. \rho^I_{ge}(t) \int_0^\infty dt^\prime e^{i \omega_0 t^\prime} \xi(t,t^\prime)  \right),  \label{AEq:C33}
\end{eqnarray}
where $\rho^I_{eg}(t)= \langle e \vert \hat{\rho}^I(t) \vert g \rangle$ and the term 
$\xi(t,t^\prime)=\text{Tr}_B \left\lbrace  \hat{B}^I(t)\hat{B}^I(t^\prime) \hat{\rho}_B^I(0) \right\rbrace+\text{Tr}_B \left\lbrace  \hat{B}^I(t^\prime)\hat{B}^I(t) \hat{\rho}_B^I(0) \right\rbrace$. Neglecting the Lamb shift, we get
\begin{eqnarray}
\Xi (t) &=&-{i}{\hbar}^2 \omega_0 \dot{\lambda}(t) \left( \rho^I_{eg}(t)  e^{-i \omega_0 t} - \rho^I_{ge}(t)  e^{i \omega_0 t}\right) \left( \Gamma_{\uparrow}+\Gamma_{\downarrow}\right) \nonumber \\
&=&{2}{\hbar}^2 \omega_0 \dot{\lambda}(t) \text{Im}\left\lbrace \rho^I_{eg}(t)  e^{-i \omega_0 t}\right\rbrace \left( \Gamma_{\uparrow}+\Gamma_{\downarrow}\right)  \label{AEq:C34} \\
&=&{2}{\hbar}^2 \omega_0 \dot{\lambda}(t) \text{Im}\left\lbrace \rho_{eg}(t) \right\rbrace \left( \Gamma_{\uparrow}+\Gamma_{\downarrow}\right).
\end{eqnarray}
With this form of $\Xi (t)$, $\langle W^3 \rangle_{S+B}$ reduces to
\begin{eqnarray}
\langle W^3 \rangle_{S+B} &\approx&\frac{\hbar^2 \omega_0}{2}  \left( \Gamma_{\uparrow}+\Gamma_{\downarrow}\right)  \int_0^\tau dt \dot{\lambda}(t) \text{Im}\left\lbrace \rho_{eg}(t) \right\rbrace. \nonumber \\ \label{AEq:SB}
\end{eqnarray}

\bibliographystyle{apsrev}
\bibliography{moments.bib}

\begin{thebibliography}{41}
\bibitem[{foo({\natexlab{a}})}]{footnote}
\bibinfo{note}{$P[E_\tau,E_0]$ can also be written as $P[E_\tau,E_0]= \vert
  \langle E_\tau \vert \hat{U}(\tau,0) \vert E_0\rangle \vert^2 \langle E_0
  \vert \rho_0 \vert E_0\rangle$. However, in the main text we use the trace
  form as it is useful in the following calculations.}
\bibitem[{foo({\natexlab{b}})}]{footnote2}
\bibinfo{note}{Note that if the observable $\hat{A}(t)=\hat{H}(t)$, then $e^{(i
  \lambda \hat{A}(t))} \hat{H}(t) e^{(-i \lambda \hat{A}(t))}=\hat{H}(t)$ in
  Eq. (20) in Ref. 24.}
\end{thebibliography}
\end{document}